# Correlation between linear resistivity and $T_c$ in organic and pnictide superconductors

Nicolas Doiron-Leyraud [1,*], P. Auban-Senzier [2], S. René de Cotret [1], A. Sedeki [1], C. Bourbonnais [1,3], D. Jérome [2,3], K. Bechgaard [4] & Louis Taillefer [1,3,*]

[1] *Département de physique and RQMP, Université de Sherbrooke, Sherbrooke, Québec J1K 2R1, Canada*

[2] *Laboratoire de Physique des Solides, UMR 8502 CNRS Université Paris-Sud, 91405 Orsay, France*

[3] *Canadian Institute for Advanced Research, Toronto, Ontario M5G 1Z8, Canada*

[4] *Department of Chemistry, H.C. Ørsted Institute, Copenhagen, Denmark*

* E-mail: ndl@physique.usherbrooke.ca, louis.taillefer@physique.usherbrooke.ca

**A linear temperature dependence of the electrical resistivity as $T \rightarrow 0$ is the hallmark of quantum criticality in heavy-fermion metals and the archetypal normal-state property of high-$T_c$ superconductors, yet in both cases it remains unexplained. We report a linear resistivity on the border of spin-density-wave order in the organic superconductor $(TMTSF)_2X$ ($X = PF_6$, $ClO_4$), whose strength scales with the superconducting temperature $T_c$. This scaling, also present in the pnictide superconductors, reveals an intimate connection between linear-$T$ scattering and pairing, shown by renormalization group theory to arise from antiferromagnetic fluctuations, enhanced by the interference of superconducting correlations. Our results suggest that linear resistivity in general may be a consequence of such interference and pairing in overdoped high-$T_c$ cuprates is driven by antiferromagnetic fluctuations, as in organic and pnictide superconductors.**

A distinctive feature of high-temperature cuprate superconductors is the linear temperature dependence of their electrical resistivity $\rho(T)$, observed most clearly near optimal doping (*1*,*2*,*3*). Indeed, above a temperature $T_\rho$ (Fig. 1D), often used as a definition of the pseudogap temperature $T^*$, $\rho = \rho_0 + AT$ (*3*,*4*). In the hole-doped cuprate $La_{1.6-x}Nd_{0.4-x}Sr_xCuO_4$ (Nd-LSCO), the resistivity at the critical hole concentration (doping, $p$) where spin/charge "stripe" order vanishes (*5*) ($p \approx 0.24$; Fig. 1D) is linear down to the lowest temperatures (Fig. 2E), when superconductivity is suppressed with a magnetic field (*6*). The same is true in the electron-doped cuprate $Pr_{2-x}Ce_xCuO_4$ just beyond its antiferromagnetic quantum critical point (*7*). However, while a linear resistivity is a universal property of cuprates, with the same $A$ coefficient per $CuO_2$ plane in different hole-doped materials (fig. S1) (*8*), in most of these materials neither antiferromagnetic nor "stripe" order has been observed near optimal doping. As a result, the origin of the linear-$T$ resistivity remains a mystery, as enigmatic as the pseudogap phase itself. In heavy-fermion metals, a linear-$T$ (or nearly linear-$T$) resistivity as $T \rightarrow 0$ has been observed at the quantum critical point where antiferromagnetic order vanishes (*9*), in some cases giving way to superconductivity (*10*). Here, the mechanism of linear-$T$ resistivity remains a subject of debate because of the uncertainty on how to treat the local *f*-electron moments through the quantum critical point (*9*). As a result, although it is remarkably simple in appearance, the purely linear dependence of $\rho(T)$ in cuprates and heavy-fermion metals has emerged as a major outstanding puzzle.

To shed light on this puzzle, we turn to a different family of materials: the Bechgaard salts $(TMTSF)_2X$ (*11*). These quasi-1D organic superconductors are simpler than the cuprates, as they have neither pseudogap phase nor Mott insulating state, and they have the following advantages: single crystals are stoichiometric and of very high purity; superconductivity is suppressed with a small magnetic field, giving easy access to the $T \rightarrow 0$ regime; tuning is done by applying pressure as opposed to chemical doping. This offers the possibility to tune, using modest pressures, a single stoichiometric sample across large sections of the phase diagram, without altering its composition, dimensions, and electrical contacts, thereby minimizing relative uncertainties and disorder-related effects. The Bechgaard salts are also simpler than the heavy-fermion metals, having a simple Fermi surface and no *f*-electron moments. Moreover, because of their quasi-1D character the behaviour of electrons in these Bechgaard salts is reliably described by a renormalization group (RG) approach that treats antiferromagnetic and superconducting correlations self-consistently (*12*). With minimal assumptions, this weak-coupling theory can account in detail (*13*) for the experimental phase diagram (Fig. 1A) and for the magnetic fluctuations measured by nuclear magnetic resonance (NMR) (*14,15*).

Here we report measurements of the *a*-axis electrical resistivity in $(TMTSF)_2PF_6$ and $(TMTSF)_2ClO_4$ down to low temperature as a function of pressure. In $(TMTSF)_2X$, the organic molecules are stacked along the *a*-axis, which gives the best conduction, followed by the intra-layer conduction between stacks (*b*-axis), and by the conduction between layers of stacks (*c*-axis), each differing by two orders of magnitude in conductivity. Below 100 K, conduction is coherent along both the *a* and *b* directions and the electron system is in a quasi-2D regime. The phase diagram of $(TMTSF)_2PF_6$ as a function of pressure shows a superconducting dome peaking at the critical point where a spin-density-wave (SDW) phase vanishes (Fig. 1A). Replacing the $PF_6$ anion by $ClO_4$ creates a high-pressure analogue that is superconducting at ambient pressure (Fig. 3A). By measuring both, we were able to cover the full phase diagram, from SDW to superconductivity to non-superconducting metal. Our two main experimental findings are: 1) a strict linear temperature dependence of $\rho(T)$ as $T \rightarrow 0$ on the border of antiferromagnetic order; 2) a simultaneous suppression of the linear resistivity and $T_c$ with increasing pressure. The scaling of linear resistivity with $T_c$ reveals an intimate link between scattering and pairing. Our main theoretical finding is a linear-$T$ scattering rate from antiferromagnetic fluctuations that are enhanced by the constructive interference of pairing correlations.

In Fig. 2A, we show the *a*-axis electrical resistivity of $(TMTSF)_2PF_6$ at $P = 11.8$ kbar, just above the critical pressure where SDW order vanishes (*16,17*); it is strictly linear from an upper temperature $T_0 \approx 8$ K down to the lowest measured temperature ($\approx 0.1$ K), reached by suppressing superconductivity with a magnetic field. This linearity of $\rho(T)$ over two decades of temperature is our first main finding. In Fig. 3A, we plot the superconducting transition temperature $T_c$ of $(TMTSF)_2ClO_4$ as a function of pressure (fig. S9) (*8*); it decreases monotonically and vanishes at $P = P_c \approx 8$ kbar. In fig. S2A (*8*), we show a typical resistivity trace for $(TMTSF)_2ClO_4$ as a function of temperature, measured at $P = 4.9$ kbar $< P_c$; the data for all other pressures are given in fig. S6 (*8*). Below 10 K, the data are best described by a fit of the form $\rho = \rho_0 + AT + BT^2$. The magnitude of the $A$ coefficient decreases monotonically with pressure and vanishes at $P_c$, the critical pressure where $T_c \rightarrow 0$ (Fig. 3A). The same analysis applied to $(TMTSF)_2PF_6$ (fig. S6) (*8*) also yields a parallel decrease of $A$ and $T_c$ (fig. S3) (*8*). In this case, $P_c$ is above the maximum pressure we could apply ($\approx 21$ kbar). When $P$ exceeds $P_c$ (in $(TMTSF)_2ClO_4$), $\rho_a(T)$ becomes purely quadratic, so that $\rho = \rho_0 + BT^2$ (Fig. 2B); in other words $A = 0$ when $T_c = 0$. This brings us to our second main finding, highlighted in Fig. 4A: $A$ scales with $T_c$ in both $(TMTSF)_2ClO_4$ and $(TMTSF)_2PF_6$.

A weak-coupling solution to the problem of the interplay between antiferromagnetism and superconductivity in the Bechgaard salts has been worked out using the functional renormalization group approach (*13*,*18*). The calculated phase diagram (Fig. 1B) captures the key features of the experimentally-determined phase diagram (Fig. 1A). The superconducting state below $T_c$ has *d*-wave symmetry (*15*), with pairing mediated by antiferromagnetic spin fluctuations. The normal state above $T_c$ is characterized by the constructive interference of antiferromagnetic and pairing correlations, which enhances the amplitude of spin fluctuations (*13*,*18*). The antiferromagnetic correlation length $\xi(T)$ increases according to $\xi = c(T + \Theta)^{-1/2}$ as $T \rightarrow T_c$, where $\Theta$ is a positive temperature scale (*13*). This correlation length is expected to impart an anomalous temperature dependence to any quantity that depends on spin fluctuations. For instance, it was shown (*13*) to account in detail for the NMR relaxation rate measured (*14,15*) in the Bechgaard salts. Through Umklapp scattering, antiferromagnetic spin fluctuations will also convey an anomalous temperature dependence to the quasi-particle scattering rate $\tau^{-1}$, in addition to the regular Fermi-liquid component which goes as $T^2$. Evaluation of the imaginary part of the one-particle self-energy yields $\tau^{-1} = a\, T\xi + b\, T^2$ (*8*), where $a$ and $b$ are constants. It is then natural to expect the resistivity to contain a linear term $AT$ (in the limit $T \ll \Theta$), whose magnitude would presumably be correlated with $T_c$, as both scattering and pairing are caused by the same antiferromagnetic correlations. Calculations of the conductivity are needed to see whether the combined effect of pairing and antiferromagnetic correlations conspires to produce the remarkably linear resistivity observed in $(TMTSF)_2PF_6$ on the border of SDW order.

Comparison with recent data on the pnictide superconductor $Ba(Fe_{1-x}Co_x)_2As_2$ (*19,20*) suggests that the salient features of electron behaviour in the organic superconductors may be a general property of metals near a SDW instability. First, the phase diagram of $Ba(Fe_{1-x}Co_x)_2As_2$ (Fig. 1C) is strikingly similar to that of $(TMTSF)_2PF_6$ (Fig. 1A), with the characteristic temperature scales ($T_{SDW}$ and $T_c$) enhanced by a factor 20. Secondly, near the critical doping where SDW order vanishes (at $x \approx 0.08$), the resistivity of $Ba(Fe_{1-x}Co_x)_2As_2$ is purely linear below $T_0 \approx 125$ K, at least down to $T_c \approx 25$ K (Fig. 2C). Thirdly, throughout the overdoped regime ($x > 0.08$), $\rho(T)$ is well described by $\rho_0 + AT + BT^2$ (fig. S2B) (*8*), with a linear term $A$ that decreases monotonically as $T_c$ drops (Figs. 3B and 4B, and fig. S4 (*8*)), vanishing at the critical doping $x_c \approx 0.18$ where $T_c \rightarrow 0$ (Fig. 1C). Finally, for $x = x_c$ and beyond, $\rho(T)$ is purely quadratic down to the lowest temperatures (Fig. 2D), so that $A = 0$ when $T_c = 0$. All this reveals a detailed similarity with $(TMTSF)_2X$ which strongly suggests that the same underlying mechanisms are at play in the pnictides and in the Bechgaard salts.

In the cuprates, it has long been known that the resistivity of strongly-overdoped non-superconducting samples (with $T_c = 0$) is quadratic at low temperature, with $\rho = \rho_0 + BT^2$, as in $Tl_2Ba_2CuO_{6+\delta}$ (Tl-2201) at $p = 0.27$ (*2*) and $La_{2-x}Sr_xCuO_4$ (LSCO) at $p = 0.30$ (*21*) (Fig. 2F). The evolution of $\rho(T)$ from $\rho_0 + AT$ near optimal doping to $\rho_0 + BT^2$ at high doping can best be described by the approximate form $\rho_0 + AT + BT^2$ at intermediate doping (*22,23,24*). For example, while the resistivity of Nd-LSCO at $p = 0.24$ is purely linear below a temperature $T_0 \approx 80$ K (Fig. 2E), the data from $T_c$ to 300 K are well described by a fit to $\rho_0 + AT + BT^2$ over the full range (fig. S2C) (*8*). Fitting published data on Nd-LSCO (*5*,*6*) and LSCO (*4*,*21*) to this form (*8*) yields the plot of $A$ vs $p$ shown in Fig. 3C, where $A$ is seen to extrapolate to zero at the same critical doping as $T_c$ does, namely $p_c \approx 0.27$. This correlation between $A$ and $T_c$ was emphasized in a recent report of data on strongly-overdoped LSCO taken in large magnetic fields (*24*). The correlation between $A$ and $T_c$ for Tl-2201 is displayed in Fig. 4C, based on fitting published data (*2*,*22*,*25*) to the same form (*8*).

A renormalization group approach (*26*) similar to that applied to the Bechgaard salts has been used to show that pairing in both pnictides (*27*) and overdoped cuprates (*26,27*) is driven by antiferromagnetic correlations, giving an out-of-phase *s*-wave symmetry in the multi-band pnictides and *d*-wave symmetry in the single-band cuprates. It also yields a linear-$T$ scattering rate in overdoped cuprates (*28*), whose strength decreases with doping and whose anisotropy agrees qualitatively with that extracted from angle-dependent magneto-resistance measurements (*29,30*).

In summary, measurements on the organic salt $(TMTSF)_2X$ show that a linear resistivity as $T \rightarrow 0$ is a property of metals that transcends the peculiarities of *f*-electron metals and copper oxides. Its occurrence does not require Kondo or Mott mechanisms. The very similar phase diagram, detailed temperature dependence of the resistivity and correlation with $T_c$ observed in the pnictide $Ba(Fe_{1-x}Co_x)_2As_2$ strongly suggest that the fundamental ingredient is a proximity to SDW order. RG calculations applied to $(TMTSF)_2X$ show that antiferromagnetic fluctuations associated with the SDW order do yield a linear-$T$ scattering rate, enhanced by the constructive interference of pairing correlations. This interference may be of fundamental importance for a general theory of the linear resistivity. The fact that the strength of the linear resistivity scales with $T_c$ in $(TMTSF)_2X$ as a function of pressure and in $Ba(Fe_{1-x}Co_x)_2As_2$ as a function of doping strongly suggests that linear-$T$ scattering and pairing have a common origin. This is empirical evidence that pairing in the Bechgaard salts and pnictides is driven by antiferromagnetic fluctuations, in agreement with the RG studies. Even though the situation in cuprates is more ambiguous, in particular because of the ill-understood pseudogap phase, the same correlation between linear resistivity and $T_c$ observed outside the pseudogap phase in several cuprates would seem to also favour, by analogy, a scenario of pairing by antiferromagnetic spin fluctuations, at least in the overdoped regime. This would be consistent with the parallel suppression of superconductivity and spin fluctuations measured by inelastic neutron scattering in overdoped LSCO (*31*) and obtained in numerical simulations of the Hubbard model (*32*).

**Acknowledgements** We thank R. Homier, F. Laliberté and H. Shakeripour for their help with the data analysis and H.H. Wen for providing us with the data in ref. 20. LT acknowledges support from the Canadian Institute for Advanced Research and funding from NSERC, FQRNT, CFI and a Canada Research Chair.

**Fig. 1.** Phase diagrams. (**A**) Temperature-pressure phase diagram of $(TMTSF)_2PF_6$, a member of the Bechgaard salt series of organic conductors (*12*), showing a spin-density-wave (SDW) phase below $T_{SDW}$ (grey circles) and superconductivity (SC) below $T_c$ (black circles) (from refs. 16 and 17, and this work). The latter phase ends at the critical pressure $P_c$. (The phase diagram of $(TMTSF)_2ClO_4$ is shown in Fig. 3A.) (**B**) Theoretical phase diagram of the quasi-1D electron gas model described in the text (and in ref. 13), showing the spin-density-wave and superconducting phases. The transverse second-nearest-neighbour hopping parameter *t'* parametrizes the change in nesting of the open Fermi surface, which simulates the effect of pressure (*13,18*). (**C**) Temperature-doping phase diagram of the pnictide superconductor $Ba(Fe_{1-x}Co_x)_2As_2$ as a function of nominal Co concentration *x* (from ref. 20), showing a metallic SDW phase below $T_{SDW}$ and superconductivity below a $T_c$ which ends at the critical doping $x_c$. (**D**) Temperature-doping phase diagram of the cuprate superconductor $La_{2-y-x}Nd_ySr_xCuO_4$, showing the temperature $T_{SDW}$ for the onset of "spin stripe" order, a type of spin-density-wave order seen by neutron diffraction (grey dots, from ref. 5), the temperature $T_\rho$ below which the resistivity deviates from its linear behaviour at high temperature (open circles, from ref. 6), and the superconducting $T_c$ (black dots, from ref. 6), which goes to zero at the critical doping $p_c$. In panels **A**, **C** and **D**, the vertical dashed line separates a regime where the resistivity $\rho(T)$ grows as $T^2$ (on the right) from a regime where $\rho(T)$ grows as $T + T^2$ (on the left) (see text).

**Fig. 2.** Temperature dependence of the resistivity, in organic (*a*-axis), pnictide (*ab*-plane) and cuprate (*ab*-plane) superconductors. *Left panels*: near their respective quantum critical point, plotted as $\rho(T)$ vs $T$: (**A**) $(TMTSF)_2PF_6$ near its SDW phase, at $P$ = 11.8 kbar; (**C**) $Ba(Fe_{1-x}Co_x)_2As_2$ near its SDW phase, at $x$ = 0.10 (from ref. 20); (**E**) Nd-LSCO near its "spin stripe" phase, at $p$ = 0.24

(from ref. 6). In panels **A** and **E**, the data in dark green (at low temperature) are in a magnetic field large enough to suppress superconductivity (30 mT and 33 T, respectively). In all cases, the resistivity is strictly linear below a temperature $T_0$ (as indicated) down to the lowest measured temperature. *Right panels*: at or slightly beyond the superconducting critical point where $T_c \rightarrow 0$, plotted as $\Delta\rho = \rho(T) - \rho_0$ vs $T^2$: (**B**) $(TMTSF)_2ClO_4$ at $P = 10.6$ kbar $\approx 1.3\ P_c$; (**D**) $Ba(Fe_{1-x}Co_x)_2As_2$ at $x = 0.184 \approx 1.0\ x_c$ (from ref. 19); (**F**) $La_{2-x}Sr_xCuO_4$ (LSCO) at $p = 0.30 = 1.1\ p_c$ (from ref. 21) and $Tl_2Ba_2CuO_{6+\delta}$ (Tl-2201) at $p = 0.27 = p_c$ (from ref. 2). In all cases, the resistivity is purely quadratic just outside the superconducting phase.

**Fig. 3.** Linear resistivity and $T_c$ versus tuning parameter. (**A**) $A$ coefficient and $T_c$ of $(TMTSF)_2ClO_4$ as a function of pressure. $T_c$ is defined as the midpoint of the superconducting transition (*8*). The solid line is a guide to the eye. $T_c$ goes to zero at $P_c \approx 8$ kbar. $A$ is extracted from fits of the form $\rho_0 + AT + BT^2$ to the data between $T_c$ and 10 K (fig. S6) (*8*). The red dashed line is a guide to the eye. The error bar on $P$ is estimated to be ± 0.25 kbar. The error bars on $A$ are explained in (*8*). The equivalent plot for $(TMTSF)_2PF_6$ is shown in fig. S3 (*8*). (**B**) $A$ coefficient (red dots) for $Ba(Fe_{1-x}Co_x)_2As_2$ as a function of $x$, from fits of the form $\rho_0 + AT + BT^2$ to the data of ref. 20 (*8*). $T_c$ (black dots) is reproduced from ref. 20. The red dashed line is a linear fit to the $A$ data points from $x = 0.08$ to 0.15. See fig. S4 (*8*) for the corresponding analysis of data in ref. 19. (**C**) Coefficient of the linear resistivity of cuprates per $CuO_2$ plane, $A_\square = A / d$, as a function of doping $p$, for LSCO (red dots), Nd-LSCO (blue dots), YBCO (purple triangle), and Tl-2201 (green square). The data are extracted from fits of the form $\rho_0 + AT + BT^2$ (*8*) to published data (*4*,*5*,*6*,*21*). The red dashed line is a linear fit to the LSCO data points (excluding $p = 0.30$). The black dots are the corresponding $T_c$ for LSCO (from ref. 4). The black line is the formula $T_c = T_c^{max}$

$[1 - 82.6 (p - 0.16)^2]$, with $T_c^{max}$ = 37 K, so that $T_c \rightarrow 0$ at $p_c$ = 0.27. In panels (**B**) and (**C**), the ± 15 % error bar on *A* comes from the uncertainty on sample dimensions.

**Fig. 4.** Correlation between *A* and $T_c$. (**A**) *A* coefficient versus $T_c$ (midpoint) from fits of the form $\rho_0 + AT + BT^2$ to the data for $(TMTSF)_2ClO_4$ (red dots) and $(TMTSF)_2PF_6$ (blue dots) (fig. S6) (*8*). The error bars on *A* and $T_c$ are explained in (*8*). The dashed line is a fit to the form $A \sim T_c^{\gamma}$ (with $\gamma$ = 0.7) to the data of $(TMTSF)_2PF_6$. (**B**) *A* coefficient versus $T_c$ for $Ba(Fe_{1-x}Co_x)_2As_2$, from fits of the form $\rho_0 + AT + BT^2$ to data in ref. 20 (*8*). See fig. S4 (*8*) for the corresponding analysis of data in ref. 19. (**C**) *A* coefficient versus $T_c$ for the overdoped cuprate Tl-2201, from fits of the form $\rho_0 + AT + BT^2$ to published data (*8*): $T_c$ = 0, 5, 10, 48, 60 and 75 K, from ref. 2 (full squares); $T_c$ = 14 K, from ref. 22, and $T_c$ = 30 K, from ref. 25 (open squares). In panels **B** and **C**, the ± 15 % error bar on *A* comes from the uncertainty on sample dimensions. In all three panels, $T_c$ is normalized by $T_c^{max}$, its maximal value in each material, and the solid line is a linear fit to the data points for $T_c / T_c^{max} < 0.75$. In all three families of superconductors, we see that $A \rightarrow 0$ as $T_c \rightarrow 0$.

---

*Figure 1 - Doiron-Leyraud et al.*

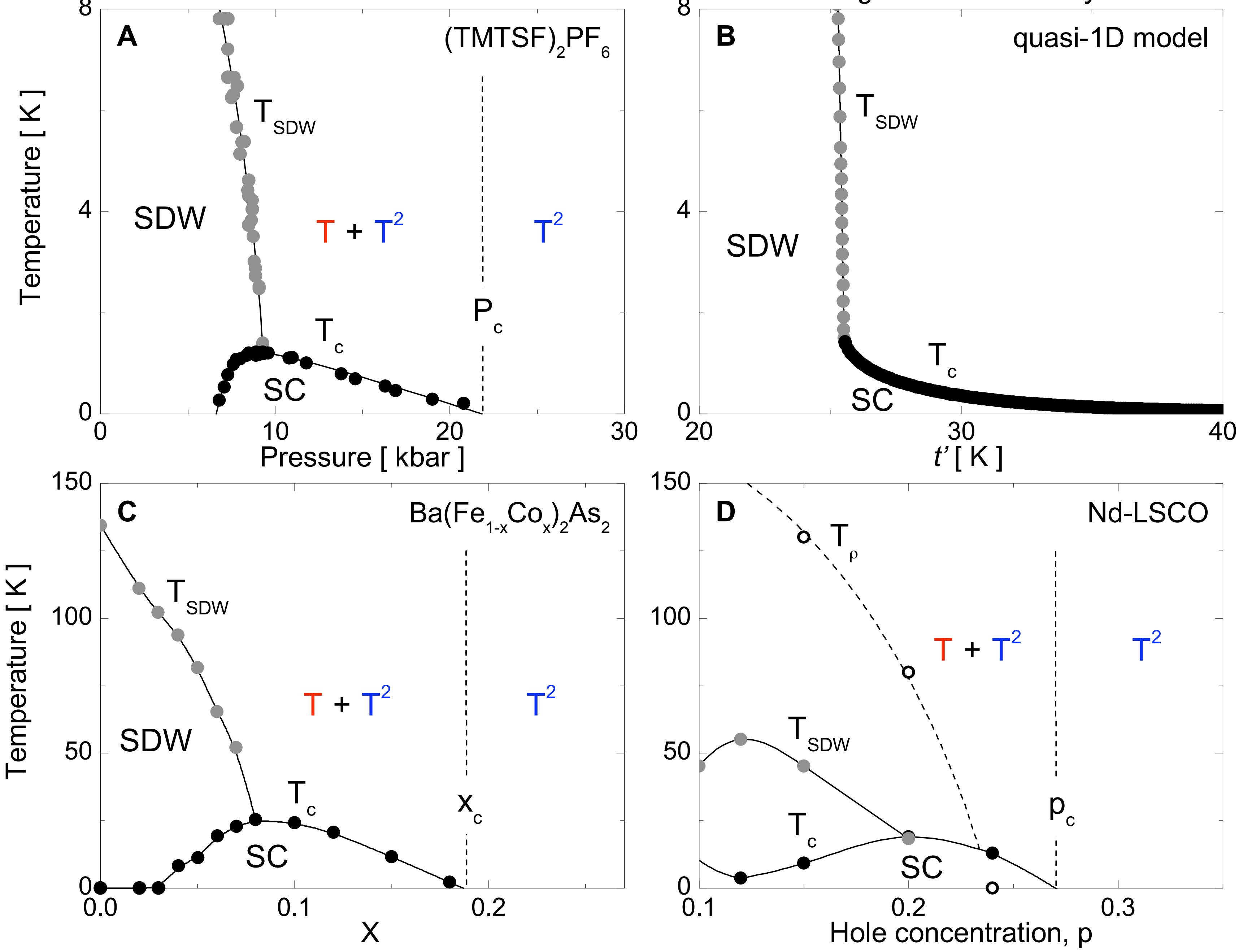

*Figure 2 - Doiron-Leyraud et al.*

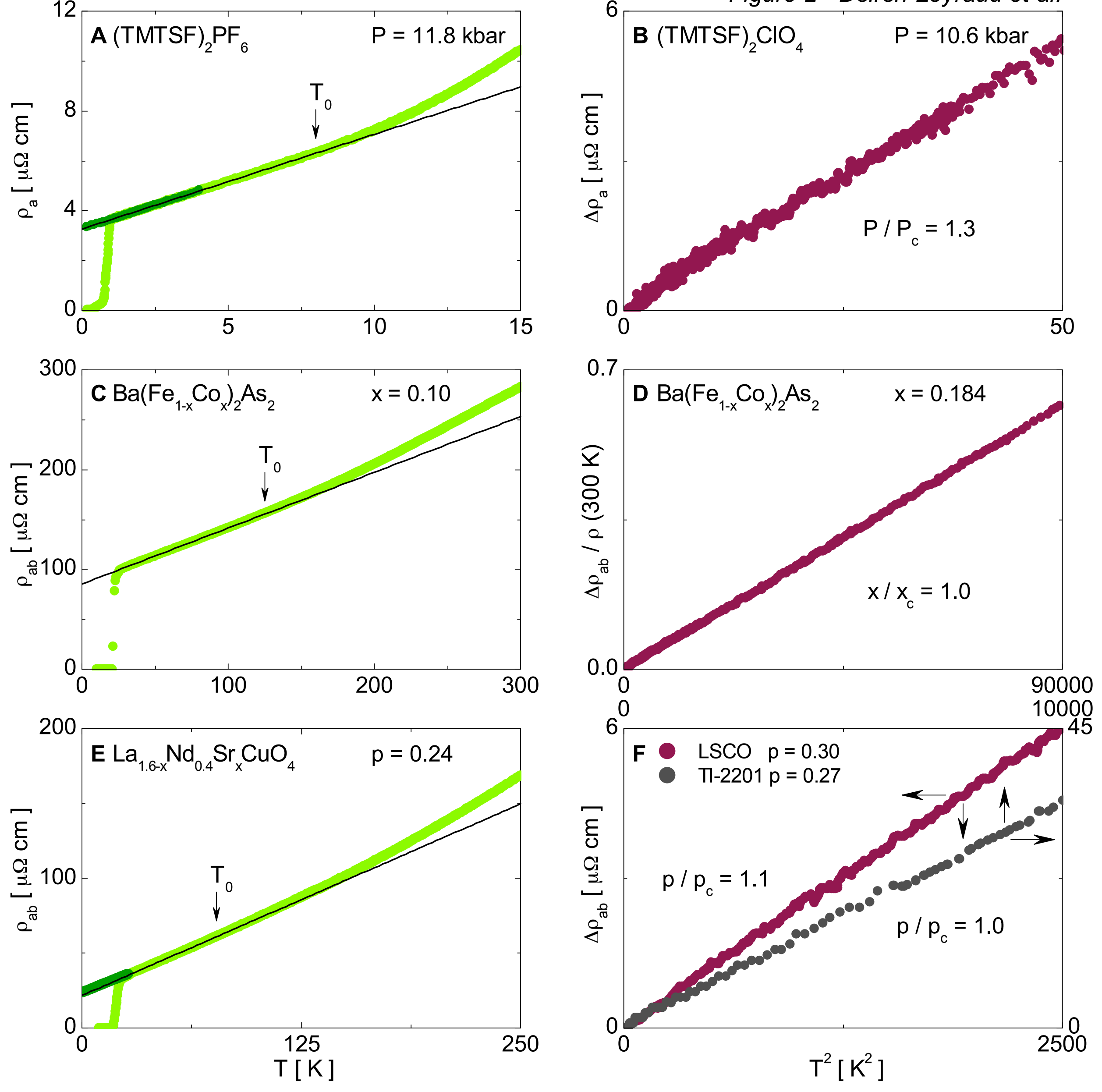

*Figure 3 - Doiron-Leyraud et al.*

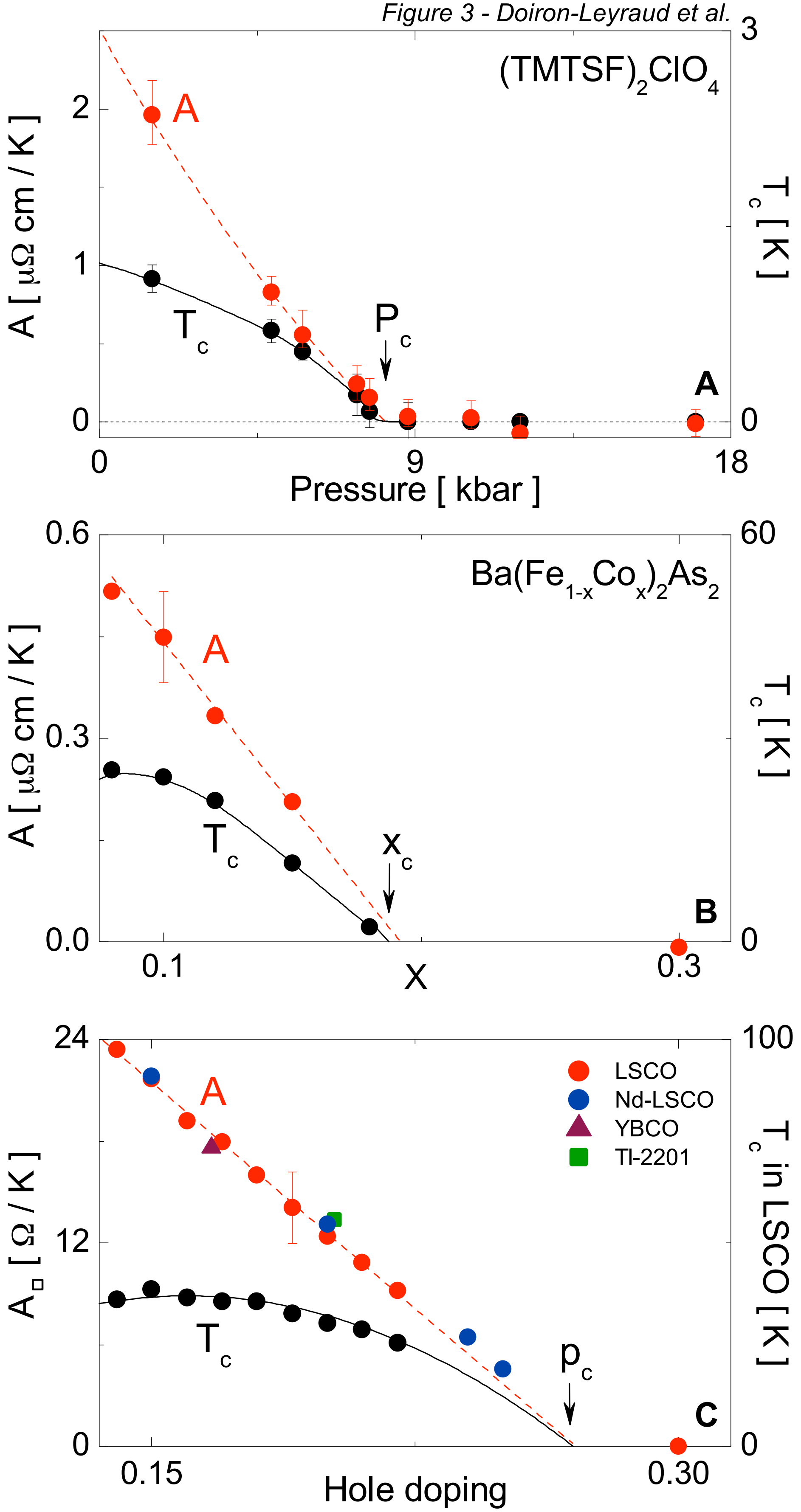

Figure 4 - Doiron-Leyraud et al.

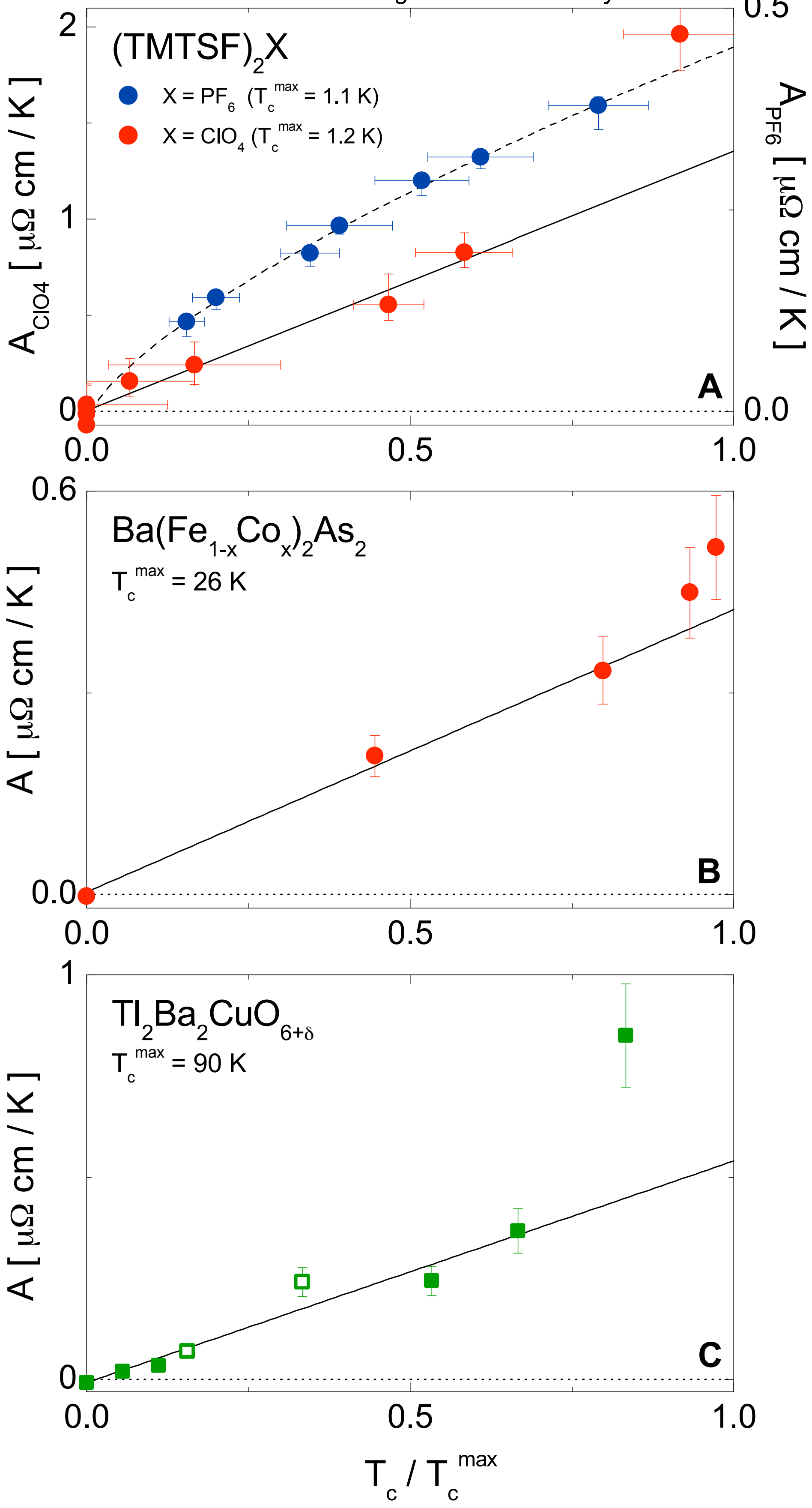

# Supporting Material for

# "Correlation between linear resistivity and $T_c$ in organic and pnictide superconductors"

Nicolas Doiron-Leyraud [1,*], P. Auban-Senzier [2], S. René de Cotret [1], A. Sedeki [1], C. Bourbonnais [1,3], D. Jérome [2,3], K. Bechgaard [4] & Louis Taillefer [1,3,*]

[1] *Département de physique and RQMP, Université de Sherbrooke, Sherbrooke, Québec J1K 2R1, Canada*

[2] *Laboratoire de Physique des Solides, UMR 8502 CNRS Université Paris-Sud, 91405 Orsay, France*

[3] *Canadian Institute for Advanced Research, Toronto, Ontario M5G 1Z8, Canada*

[4] *Department of Chemistry, H.C. Ørsted Institute, Copenhagen, Denmark*

* E-mail: ndl@physique.usherbrooke.ca, louis.taillefer@physique.usherbrooke.ca

This Supporting Material file is divided into three parts:

1) Supplementary figures S1 to S4

2) Methods (with associated fig. S9)

3) Data and Fitting (with associated figs. S5, S6, S7, S9 and S10)

## SUPPLEMENTARY FIGURES

**Fig. S1. Universal linear-*T* resistivity in cuprates.**

In-plane resistivity of four different cuprate materials, measured per $CuO_2$ plane, as $\rho_{\square} = \rho / d$, where $d$ is the average separation between $CuO_2$ planes.

(**A**) LSCO (red; from ref. 1), Nd-doped LSCO (Nd-LSCO) (blue; from ref. 2), and Tl-2201 (green; from ref. 3), all at a hole concentration (or doping) $p = 0.20$.

(**B**) LSCO (red; from ref. 4) and $YBa_2Cu_3O_y$ (YBCO) (purple; from ref. 1), both at a doping $p = 0.17$. The value of $p$ is estimated as described in the Data and Fitting section below. At these doping values, $\rho(T)$ is essentially linear above 100 K in all cases, so that $\rho_{\square}(T) \approx \rho_0 + A_{\square}T$, and the slope is the same in different materials at the same doping. In other words, $A_{\square}$ appears to have a universal magnitude and doping dependence amongst hole-doped cuprates.

**Fig. S2. Temperature dependence of the resistivity.**

(**A**) $a$-axis resistivity $\rho_a$ for the organic superconductor $(TMTSF)_2ClO_4$ at $P$ = 4.9 kbar. (**B**) In-plane electrical resistivity $\rho_{ab}$ for the pnictide $Ba(Fe_{1-x}Co_x)_2As_2$ at $x = 0.10$ (from ref. 5) and (**C**) the cuprate Nd-LSCO at a doping $p = 0.24$ (from ref. 2) The black line is a fit to the data of the form $\rho = \rho_0 + AT + BT^2$, from $T_c$ up to or 10 K (**A**) and up to 300 K (**B**, **C**).

**Fig. S3. Correlation between *A* and $T_c$ in $(TMTSF)_2PF_6$.**

(**A**) $A$ coefficient (red dots) and $T_c$ (black dots; midpoint of superconducting transition) for $(TMTSF)_2PF_6$ as a function of pressure $P$, from fits of the form $\rho_0 + AT + BT^2$ to the resistivity data (see data and fits in Figs. S6c and S6d), for *P* values above the critical point for SDW order ($P \approx 10$ kbar). (**B**) *A* coefficient versus $T_c$. The error bars are described below in the Data and Fitting section (for *A*) and Methods section (for $T_c$).

**Fig. S4. Correlation between *A* and $T_c$ in $Ba(Fe_{1-x}Co_x)_2As_2$.**
(**A**) *A* coefficient (red dots) for $Ba(Fe_{1-x}Co_x)_2As_2$ as a function of *x* for values above the SDW phase, from fits of the form $\rho_0 + AT + BT^2$ to the resistivity data of ref. 6 (see Data and Fitting section). The published data are normalised by the resistivity at 300 K. To remove effects related to changes in the residual resistivity, we normalised *A* by $\Delta\rho(300\text{ K}) = \rho(300\text{ K}) - \rho_0$. $T_c$ (black dots) is reproduced from ref. 6. Note that *x* in ref. 6 is obtained from electron microprobe analysis (EMPA), which yields slightly lower values than the nominal value from the melt, used in ref. 5. (**B**) *A* coefficient versus $T_c$. The dashed line is a guide to the eye.

## METHODS

**Crystals**. Single crystals of $(TMTSF)_2ClO_4$ and $(TMTSF)_2PF_6$ were grown by the usual method of electrocrystallization described in Refs. 7 and 8. The samples used here showed typical values of *a*-axis conductivity near 500 $(\Omega\text{ cm})^{-1}$ at room temperature and pressure. Typical sample dimensions are 1.5 x 0.2 x 0.05 $mm^3$ with the length, width and thickness along the *a*, *b*, and *c* crystallographic axes, respectively. The uncertainty on each dimension is about 0.005 mm, leading to an overall uncertainty of about ±12% on the geometric factor. The current was applied along the *a*-axis and the magnetic field along the *c*-axis. Electrical contacts were made with evaporated gold pads (typical resistance between 1 and 10 Ω) to which 17 μm gold wires were glued with silver paint.

**Measurements.** The electrical resistivity of our samples was measured with a resistance bridge using a standard four-terminal AC technique at 16 Hz. Low excitation currents of typically 30 μA were applied in order to eliminate heating effects caused by the contact resistances. This was checked using different values of current above and below this value, at temperatures below 1 K.

**Pressure.** A non-magnetic piston-cylinder pressure cell as described in Ref. 9 was employed, with Daphne oil as pressure transmitting medium. The pressure at room temperature and 4.2 K was measured using the change in resistance and superconducting $T_c$ of a Sn sample, respectively. Only the values recorded at 4.2 K are quoted here. The highest pressure we could safely achieve at low temperature was about 21 kbar, the pressure at which Daphne oil freezes at room temperature.

**Cooling.** From room temperature down to 77 K the cooling rate was kept below 1 K/min to ensure gradual freezing of the pressure medium and an optimal level of pressure homogeneity, and to avoid the appearance of cracks in our samples. While cracks were detected in some samples, all the data reported here are on samples that showed no sign of cracks, *i.e.*, their resistance at room temperature always recovered its initial value prior to each cooling cycle. No cracks were detected during pressurization either, *i.e.*, the resistance at room temperature evolved smoothly with applied pressure. Below 77 K, the cooling rate was kept below 5 K/hour to ensure adequate thermal equilibrium between the samples and the temperature sensors placed outside the cell. In the case of $(TMTSF)_2ClO_4$, slow cooling is vital to optimise anion ordering which occurs at 24 K at low pressure. The smooth evolution of both the superconducting $T_c$ (Fig. 3A) and the residual resistivity (fig. S9) with pressure, and the fact that our highest $T_c$ is comparable to the maximum $T_c$ observed in this material, all indicate that we were well into the anion-ordered, or so-called “relaxed”, state.

**Superconducting transition temperature $T_c$.** The superconducting transition temperature $T_c$ of our samples of $(TMTSF)_2ClO_4$ and $(TMTSF)_2PF_6$ was determined using the measured resistivity, as shown in fig. S9 for the case of $(TMTSF)_2ClO_4$. The $T_c$ values plotted in Figs. 3 and 4 (and S5) correspond to the midpoint of the transition, defined as the temperature where the resistance, or an extension of the initial slope, has dropped to 50% of its value at the onset of the transition. The width of the transition, shown by the error bars in Figs. 3 and 4 (and S3), correspond to the width of the resistivity drop, measured between the points where it has reached 90% and 10% of its value at the onset. The 10% criterion is determined along an extension of the intrinsic initial slope of the resistivity drop, not the foot of the transition caused by extrinsic effects (fig. S9).

**Scattering rate.** The quasi-particle scattering rate $\tau^{-1}$ at the Fermi surface is obtained from the imaginary part of the one-particle self-energy

$$\tau^{-1} = \sum_{\vec{q}} g^2(\vec{q}) \int d\omega' \left[ f(\omega') + n_B(\omega') \right] \chi''(\vec{q},\omega')\, \delta(\omega' - E(\vec{k}_F + \vec{q}))$$

where $\chi''$ is the imaginary part of the dynamic spin susceptibility at $\vec{q}$ and $\omega$; $f$ and $n_B$ are the Fermi and Bose distributions, $E(\vec{k})$ is the energy spectrum, and $g(\vec{q})$ is the Umklapp electron-electron coupling constant. From the analysis of the NMR relaxation rate data of the Bechgaard salts (*10*), spin fluctuations probed in the metallic state that precede superconductivity consist of the superposition of two contributions in momentum space. The first is of the Fermi liquid or Korringa type that dominates the relaxation rate in the high temperature part of the metallic state, and which comes from spin fluctuations at $\vec{q} = (q \sim 0, q_b)$, that is for small longitudinal momentum. The second contribution is antiferromagnetic in character and peaked at the best nesting vector

$$\vec{q} \sim \vec{q}_0 = (2k_F, \pi)$$

which leads to a strong enhancement of the relaxation rate at low temperature. Thus following the example of the NMR, the quasi-particle scattering rate superimposes both contributions, and then $\tau^{-1}$ splits into two terms and becomes $\tau^{-1} = \tau_{FL}^{-1} + \tau_{AF}^{-1}$.

For the Fermi liquid part, $\tau_{FL}^{-1}$, the ratio $\chi''(\vec{q},\omega)/\omega$ is essentially constant in the frequency range where $f + n_B$ varies with $\omega$. The integral over frequency then yields the contribution

$$\tau_{FL}^{-1} \cong \frac{\pi}{16t_\perp}\tilde{g}^2(0,\pi)\,T^2$$

which is characteristic of a Fermi liquid. Here $t_\perp$ (~ 200K) is the inter-chain hopping along the *b* direction (*11*). The transverse Umklapp electron-electron interaction $\tilde{g}(0,\pi)$ ~0.5 (normalized by the density of the states at the Fermi level), of the order of half the bandwidth, is compatible with the Fermi liquid or Korringa component of the nuclear relaxation rate in the metallic state of the Bechgaard salts (*10*) For the antiferromagnetic component, the expression of the imaginary part of the dynamic susceptibility is given by

$$\chi''(\vec{q}+\vec{q}_0,\omega) = \frac{\chi(\vec{q}_0)\Gamma\omega}{\left(1+\xi_a^2 q^2+\xi_b^2 q_b^2\right)^2+(\Gamma\omega)^2},$$

where $\xi_{a,b}$ is the antiferromagnetic correlation length along the *a* (parallel) and *b* (perpendicular) directions, $\Gamma$ is the relaxation time of spin fluctuations and $\chi(\vec{q}_0)$ is the static antiferromagnetic susceptibility. At low temperature, namely in the region where the NMR relaxation rate is strongly enhanced, $\chi(\vec{q}_0)$, and in turn $\Gamma$ are large, so that $\chi''$ is peaked at ω of the order or smaller than the temperature. Under these conditions, one finds

$$\tau_{AF}^{-1} \cong \frac{2\sqrt{2}}{\pi}\tilde{g}^2(\vec{q}_0)\,\frac{t_a}{t_\perp}T\,\xi_a,$$

where $t_a/t_\perp$ is the ratio (~ 10) between longitudinal and transverse hopping integrals[11]. The normalized longitudinal electron-electron Umklapp scattering amplitude is a small coupling; it is estimated to be $\tilde{g}(\vec{q}_0) \approx 0.02$ in the superconducting Bechgaard salts as a consequence of the weak dimerization of the organic stacks, which is essential to the existence of longitudinal Umklapp scattering at half-filling (*10,12*). At low temperature, the Curie-Weiss behaviour seen in the nuclear relaxation rate implies that $\xi_a = c(T+\Theta)^{-1/2}$ for the longitudinal antiferromagnetic correlation length. The scale $\Theta$ increases rapidly under pressure, namely as $T_c$ and the strength of antiferromagnetic fluctuations decrease. Renormalization group calculations in the framework of the quasi-one-dimensional electron gas model show that in the range of pressure of interest, we have $\xi_0 = c/\sqrt{\Theta} \approx 50...100$ as $T \rightarrow 0$ (*10*).

The total quasi-particle scattering rate can then be written in the form

$$\tau^{-1} = aT\xi_a + bT^2.$$

With the above set of figures, one gets the estimate $ac/b \sim 10 \ldots 100$ for the ratio of the coefficients associated to each contribution in the pressure range considered. As for the coefficient $a\xi_0$ of the linear temperature dependence of the scattering rate $\tau_{AF}^{-1} \approx a\xi_0 T$ at $T << \Theta$, it decreases faster than the Fermi liquid coefficient $b$ under pressure (essentially linked to the decrease of the density of states at the Fermi level).

## DATA & FITTING

**Data.** The resistivity curves for LSCO, Nd-LSCO, YBCO, Tl-2201, and $Ba(Fe_{1-x}Co_x)_2As_2$ (those from Chu *et al.*, ref. 6) were obtained by digitizing the data from the references quoted in the present document (see captions of figs. S1 and S10). We are grateful to Professor H. H. Wen for providing us with the $Ba(Fe_{1-x}Co_x)_2As_2$ data of Ref. 5. We use the doping values (hole concentration $p$ or cobalt content $x$) and $T_c$ quoted in the original references throughout. For YBCO and Tl-2201, the hole concentration $p$ is determined from the usual relation (*13*) $T_c = T_c^{max}$ [ $1 - 82.6\,(p - 0.16)^2$ ] with $T_c^{max} = 93.5$ and 90 K, respectively. The data on $(TMTSF)_2ClO_4$ and $(TMTSF)_2PF_6$ come from the present study.

**Fitting procedure**. All the data were fitted to a polynomial of the form $\rho(T) = \rho_0 + AT + BT^2$ using a 20-points fit. The temperature interval over which the data was fitted is specified in the caption of fig. S10. For LSCO, Nd-LSCO, YBCO, Tl-2201, and $Ba(Fe_{1-x}Co_x)_2As_2$, there is an error bar of about ±15% on the fitting parameters, as shown in Figs. 3B and 3C, coming from the systematic uncertainty on the sample dimensions. This error is between each sample, not between *A* and *B* for one same sample. For $(TMTSF)_2ClO_4$ and $(TMTSF)_2PF_6$ there is no error bar associated with sample dimensions between the points because in each case the data we present are on one same sample. For the latter two materials, the error bars shown in Figs. 3A, 4A, S3, S10E, and S10F come from varying the upper bound of the fit by ± 2 K.

**Fig. S5. Resistivity of the organic superconductor $(TMTSF)_2X$.**
Temperature-dependent part $\Delta\rho(T) = \rho(T) - \rho_0$ of the *a*-axis electrical resistivity of $(TMTSF)_2X$ as a function of temperature. (**A**) Data for X = $PF_6$ at pressures of 8.4, 11.8 and 20.8 kbar. The upturn at ~ 8 K for 8.4 kbar is caused by the onset of the SDW phase (see phase diagram in Fig. 1A). (**B**) Data for X = $ClO_4$ for a range of pressures going from 1.5 to 17.0 kbar.

**Fig. S6A,B. Resistivity of the organic superconductor $(TMTSF)_2ClO_4$.**
*a*-axis electrical resistivity of $(TMTSF)_2ClO_4$ as a function of temperature for a range of pressures *P*. The red lines are fits to the data according to the fitting procedure described above. The lower bound of the fitting temperature range is $T_c$. The upper bound is 10 K, determined by the limit above which the fit begins to depart significantly from the data. Changing the upper bound by ± 2 K gives the error bars on the fitting parameters *A* and *B* shown in Figs. 3A, 4A and S10E.

**Fig. S6C,D. Resistivity of the organic superconductor $(TMTSF)_2PF_6$.**
*a*-axis electrical resistivity of $(TMTSF)_2PF_6$ as a function of temperature for a range of pressures *P*. The red lines are fits to the data according to the fitting procedure described above. The lower bound of the fitting temperature range is $T_c$. The upper bound is 8 K, determined by the limit above which the fit begins to depart significantly from the data. Changing the upper bound by ± 2 K gives the error bars on the fitting parameters *A* and *B* shown in Figs. 4A, S3 and S10F.

**Fig. S7. Resistivity of the pnictide superconductor $Ba(Fe_{1-x}Co_x)_2As_2$.**
Temperature-dependent part $\Delta\rho(T) = \rho(T) - \rho_0$ of the in-plane (*ab*) electrical resistivity of $Ba(Fe_{1-x}Co_x)_2As_2$ as a function of temperature. (**A**) Data from Ref. 6 (the published data are already normalised to $\rho$(300 K)) for a range of measured cobalt content *x* going from 0.07 to 0.184. The curves are offset by 0.04 for clarity. (**B**) Data from Ref. 5 for a range of nominal cobalt content *x* going from 0.08 to 0.30.

**Fig. S9. Superconducting transition in $(TMTSF)_2X$.**
Superconducting transition in the *a*-axis electrical resistivity of (**A**) $(TMTSF)_2ClO_4$ and (**B**) $(TMTSF)_2PF_6$, for a range of pressures, as indicated. (**C**) $T_c$ defined as the midpoint of the transition, namely the temperature at which the resistivity has reached half its value at the onset of the drop. The error bar on $T_c$ is given by the 10% – 90% width of the transition.

**Fig. S10. Fit parameters *A* and *B* for organics, pnictides, and cuprates.**
Parameters *A* and *B* of our fits $\rho(T) = \rho_0 + AT + BT^2$ versus tuning parameter. (**A**) *A* and *B* versus pressure for $(TMTSF)_2ClO_4$ from fits to our data shown in Figs. S6A and S6B. The lines are guides to the eye. Here we normalise the data to $\Delta\rho(15\text{ K}) = \rho(15\text{ K}) - \rho_0$ to remove effects associated with a rapid change in electrical resistivity of this material at low pressure. While the same is true for $(TMTSF)_2PF_6$, the data are not contaminated significantly by this effect since for this material the region of interest to the current study is above 10 kbar where most of the drop in resistivity has already taken place. (**B**) *A* and *B* versus pressure for $(TMTSF)_2PF_6$, coming from fits to our data shown in Figs. S6c and S6d. The lines are guides to the eye. The error bars in **A** and **B** come from varying the upper bound of the fit range by ± 2 K. (**C**) *A* and *B* versus measured cobalt content *x* for $Ba(Fe_{1-x}Co_x)_2As_2$ from fits to the data of Ref. 6. The published data are normalised by the resistivity at 300 K and to remove effects related to the varying residual resistivity we normalised *A* and *B* by $\Delta\rho$(300 K) = $\rho$(300 K) - $\rho_0$. (**D**) *A* and *B* versus nominal cobalt content *x* for $Ba(Fe_{1-x}Co_x)_2As_2$, from fits to the data of Ref. 5. In both **C** and **D**, a constant fitting temperature range from 30 to 300 K was used to extract the fit parameters. All lines are guides to the eye. (**E**) *A* and *B* versus doping *p* for LSCO (full symbol) and Nd-LSCO (open symbols). The LSCO points for *p* going from 0.14 to 0.22 come from fits to the data of Ref. 1. The point at *p* = 0.30 comes from Ref. 14. The Nd-LSCO points at *p* = 0.15, 0.25 and *p* = 0.20, 0.24 come from fits to the data of Refs. 15 and 2, respectively. The lower bound of the fitting temperature range is determined by either 1) the departure from linearity caused by the opening of the pseudogap, 2) the range of published data available for fitting, or 3) the onset of the superconducting drop. The red dashed line is a linear fit to the LSCO data points (except the one at *p* = 0.30). The blue line is a guide to the eye. (**F**) *A* and *B* versus doping *p* for Tl-2201. The points at *p* = 0.25 and 0.26 come from fits to the data of Refs. 3 and 16, respectively. The other points come from the data of Ref. 17. The lower bound of the fitting temperature range is determined by the onset of the superconducting drop. The red line is a guide to the eye. In **E** and **F**, the upper bound of the fit range is determined by the limit above which the fit begins to depart significantly from the data. The vertical black dashed lines in all panels mark the end of the superconducting phase, so that $T_c$ = 0 to the right of each line. In the case $(TMTSF)_2PF_6$ (**B**), the actual value of Pc has not been measured. Extrapolation of $T_c$ vs *P* data (fig. S3) yields an approximate upper bound of 25 kbar, as indicated.

---

[1] Y. Ando *et al.*, *Phys. Rev. Lett.* **93**, 267001 (2004).

[2] R. Daou *et al.*, *Nature Phys.* **5**, 31 (2009).

[3] A. W. Tyler *et al.*, *Phys. Rev. B* **57**, 728 (1998).

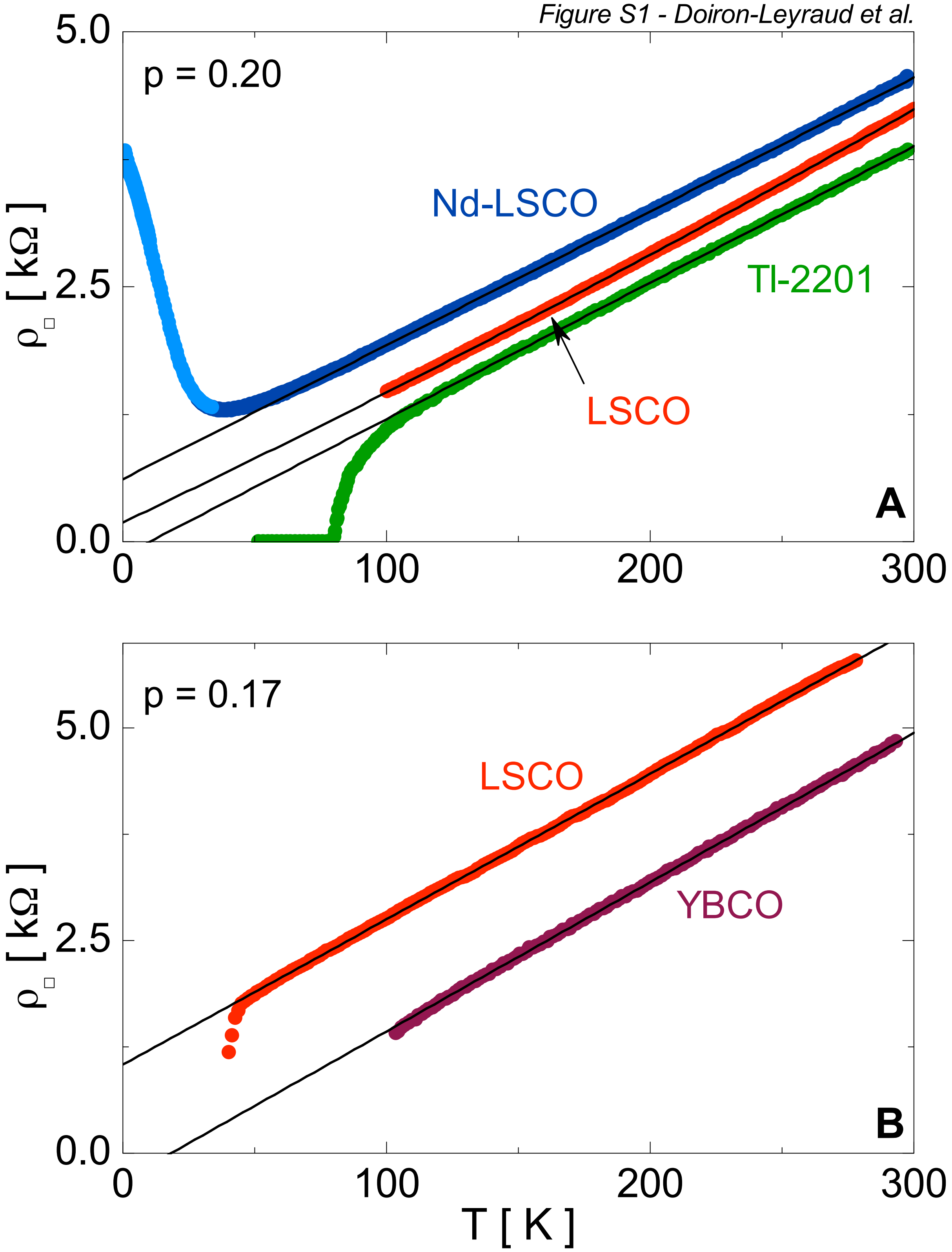

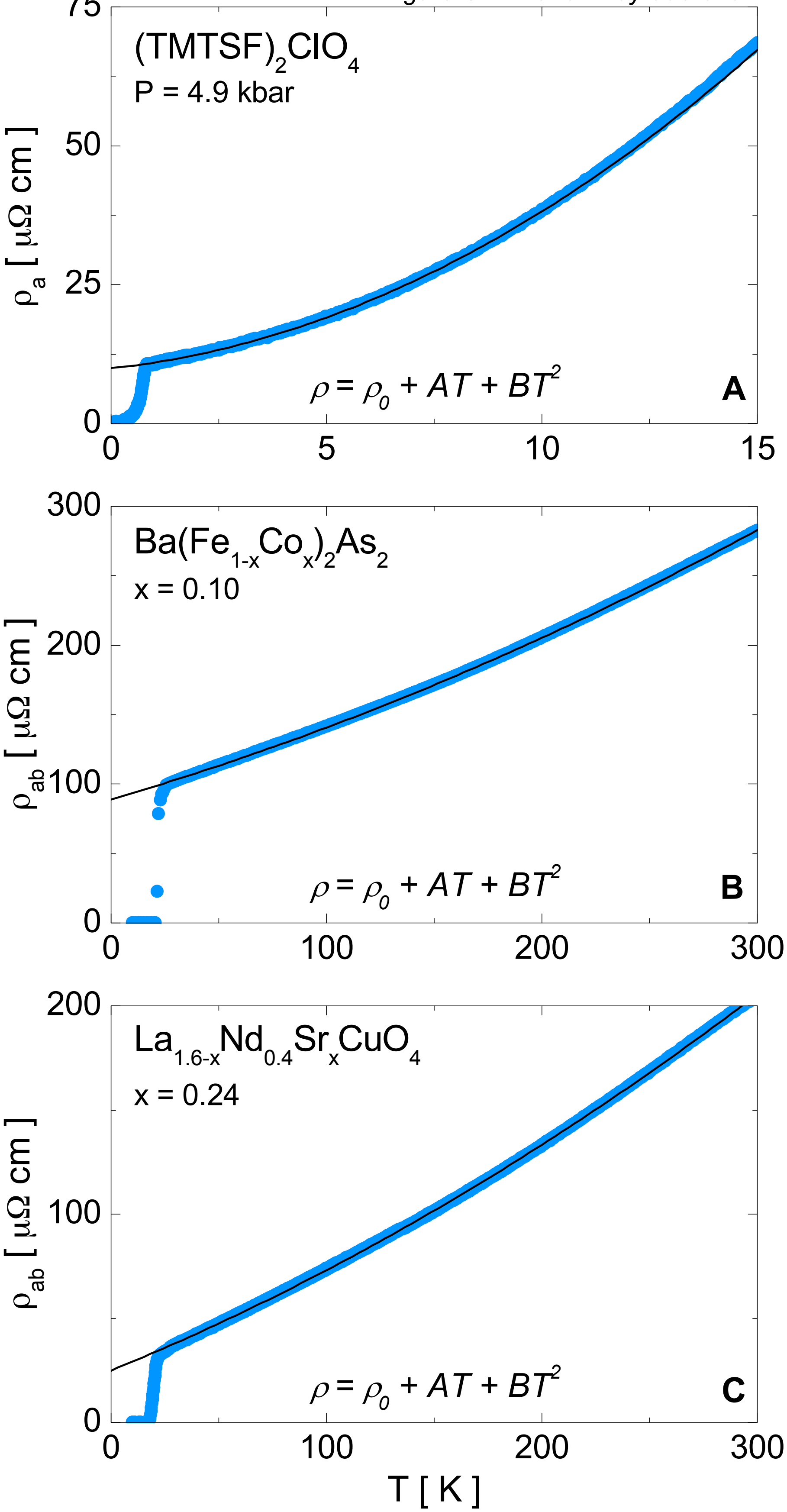
(TMTSF)2ClO4
P = 4.9 kbar
ρa [ μΩ cm ]
ρ = ρ0 + AT + BT2
A
0
25
50
75
0
5
10
15
Ba(Fe1-xCox)2As2
x = 0.10
ρab [ μΩ cm ]
ρ = ρ0 + AT + BT2
B
0
100
200
300
0
100
200
300
La1.6-xNd0.4SrxCuO4
x = 0.24
ρab [ μΩ cm ]
ρ = ρ0 + AT + BT2
C
0
100
200
0
100
200
300
T [ K ]

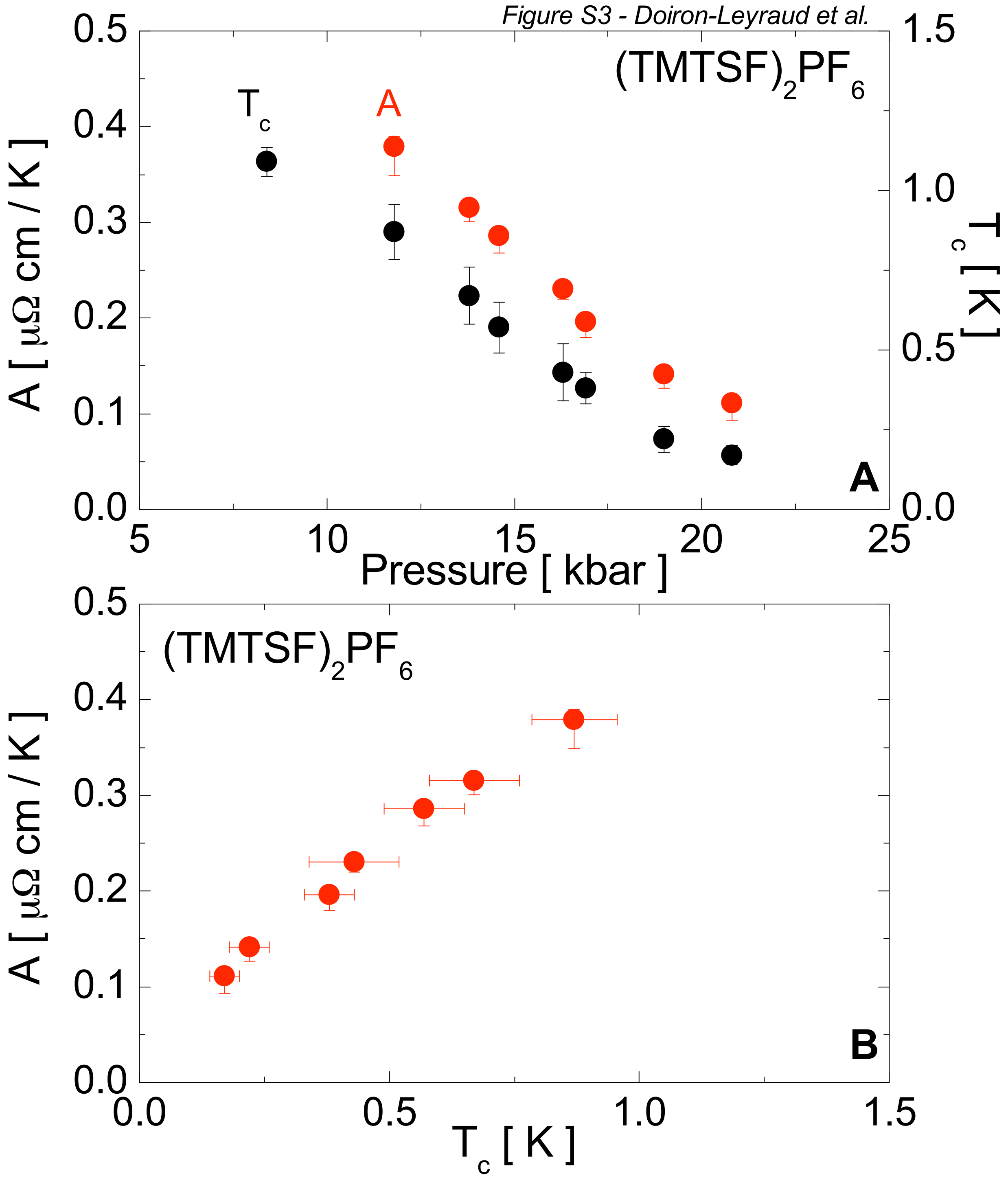

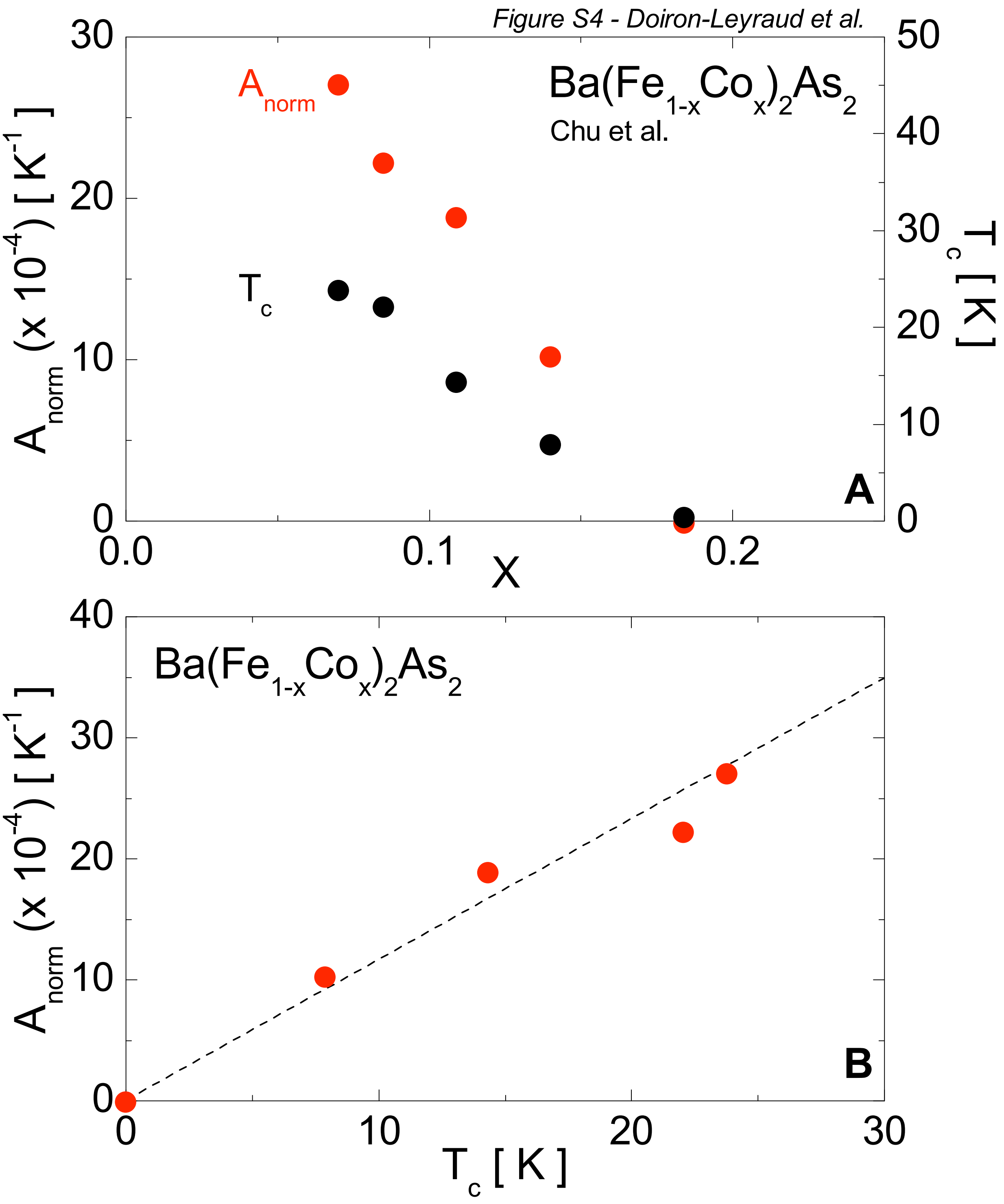

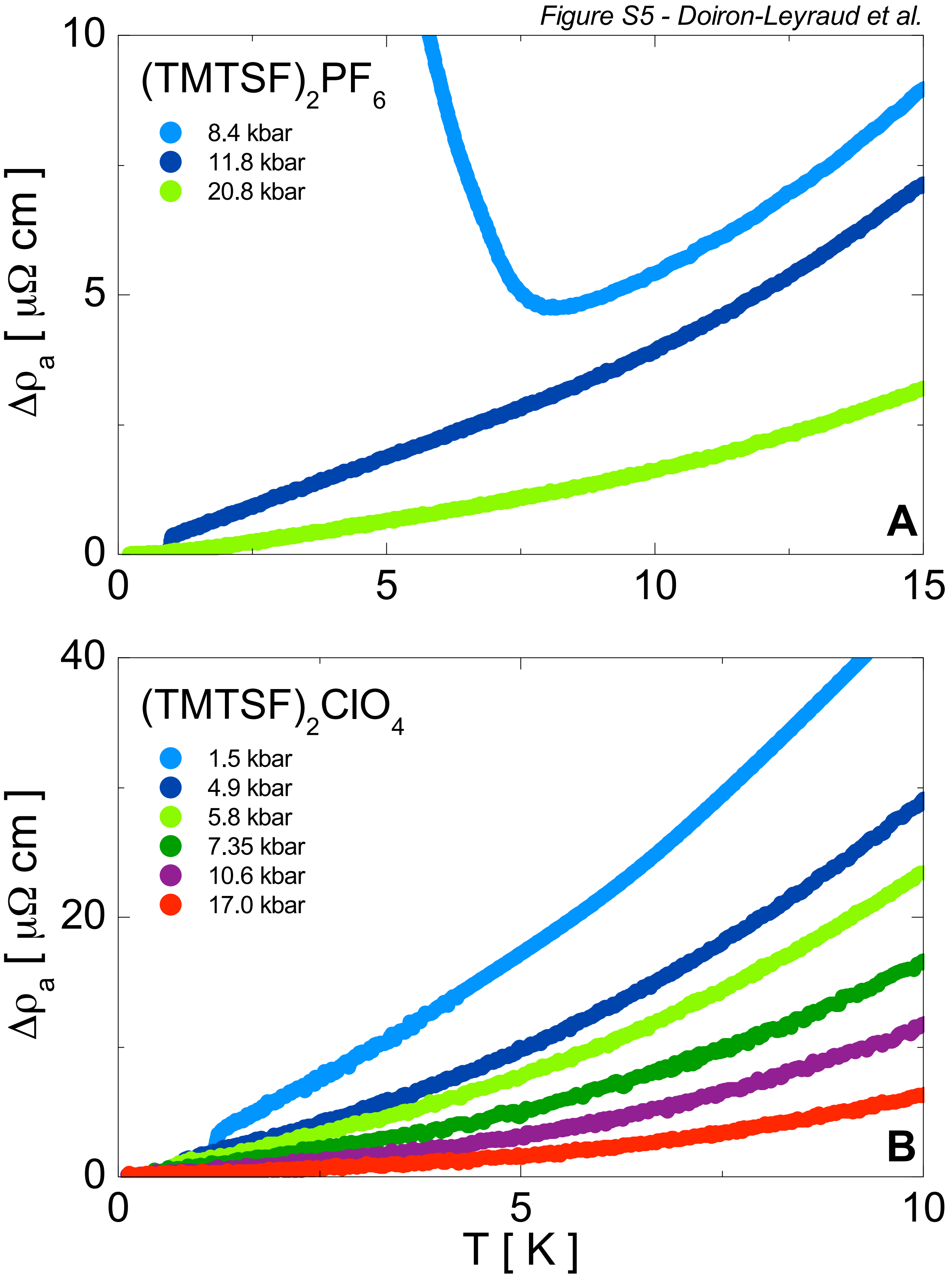
Figure S5 - Doiron-Leyraud et al.
(TMTSF)2PF6
8.4 kbar
11.8 kbar
20.8 kbar
Δρa [ μΩ cm ]
A
(TMTSF)2ClO4
1.5 kbar
4.9 kbar
5.8 kbar
7.35 kbar
10.6 kbar
17.0 kbar
Δρa [ μΩ cm ]
B
T [ K ]

*Figure S6a - Doiron-Leyraud et al.*

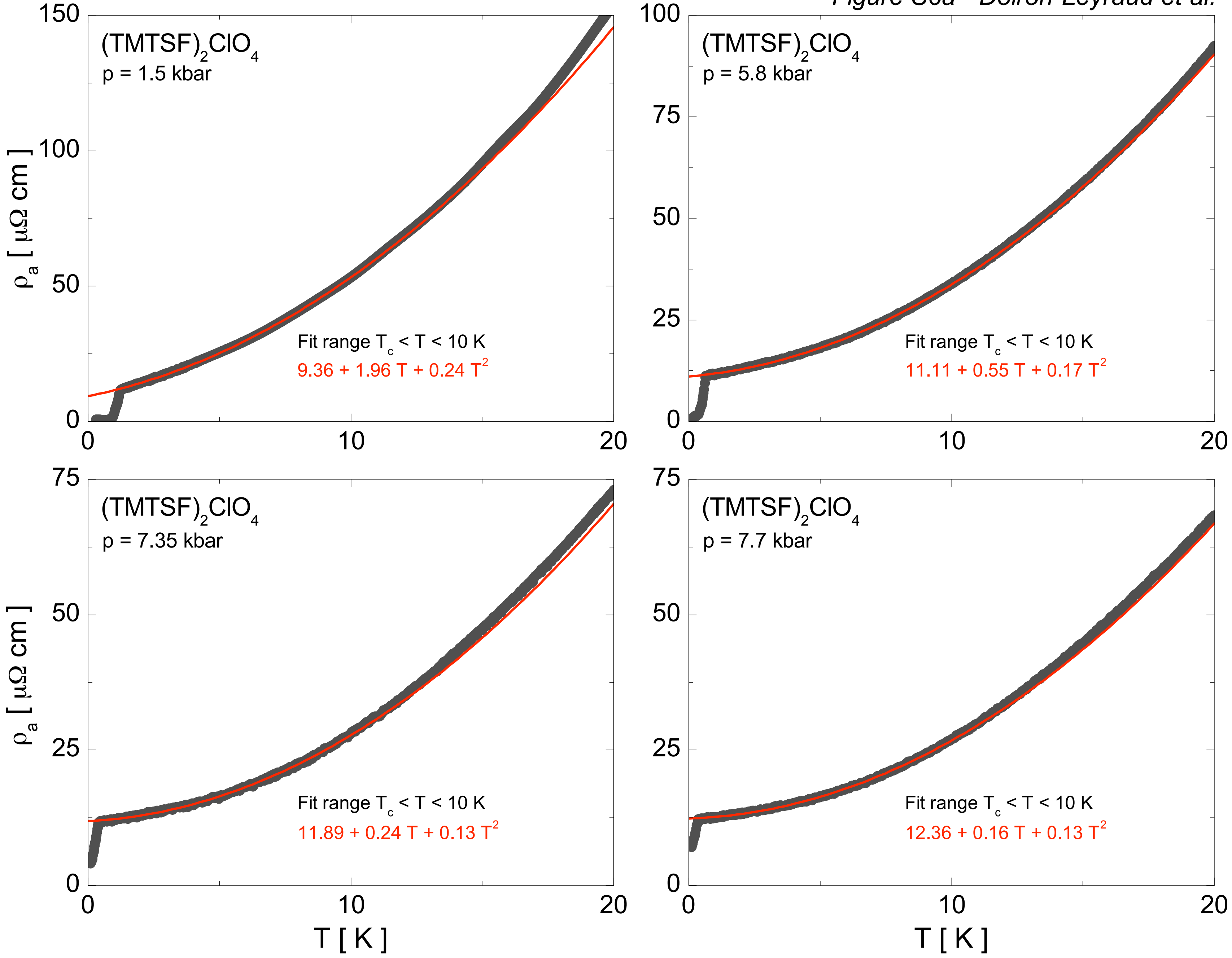

*Figure S6b - Doiron-Leyraud et al.*

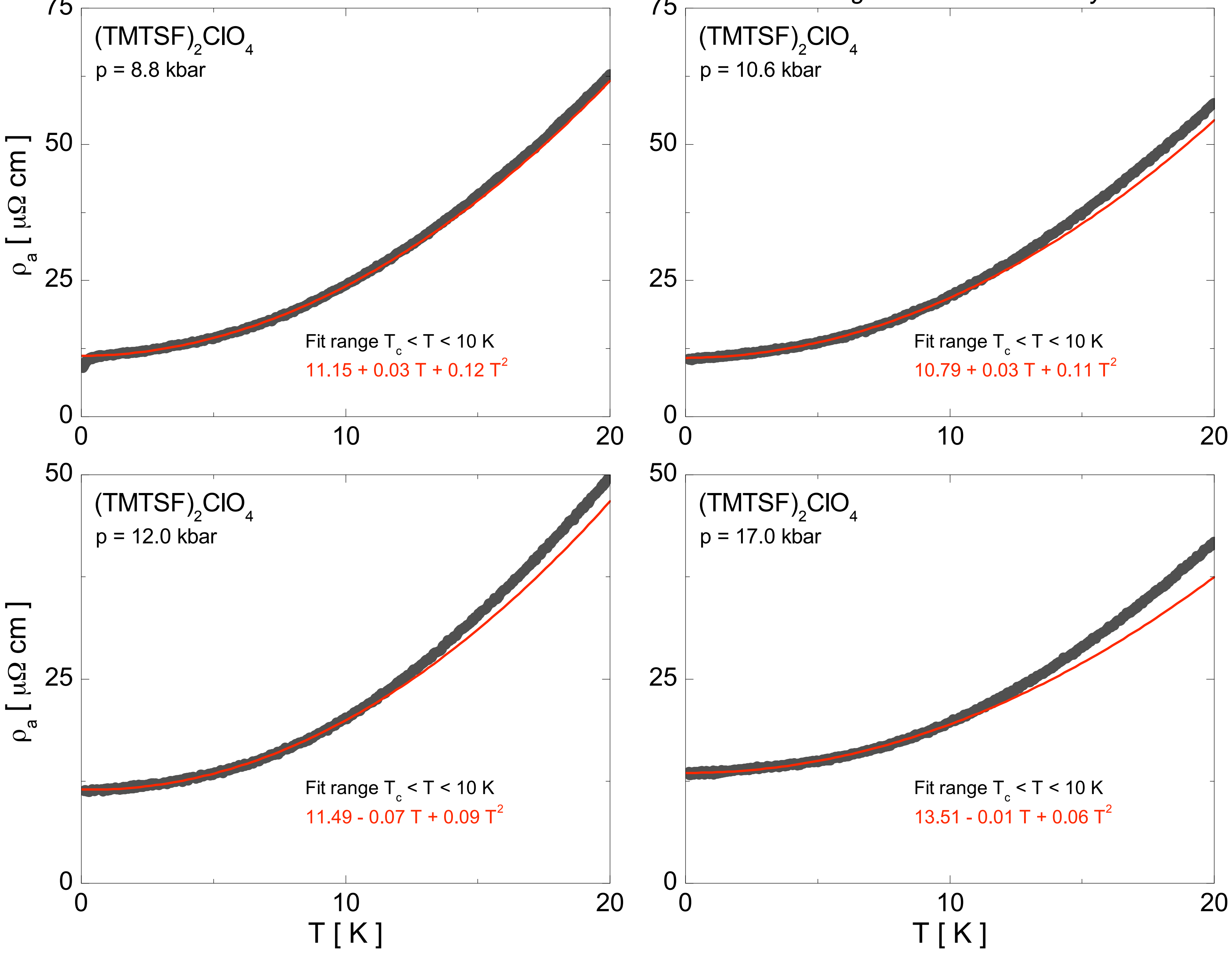

*Figure S6c - Doiron-Leyraud et al.*

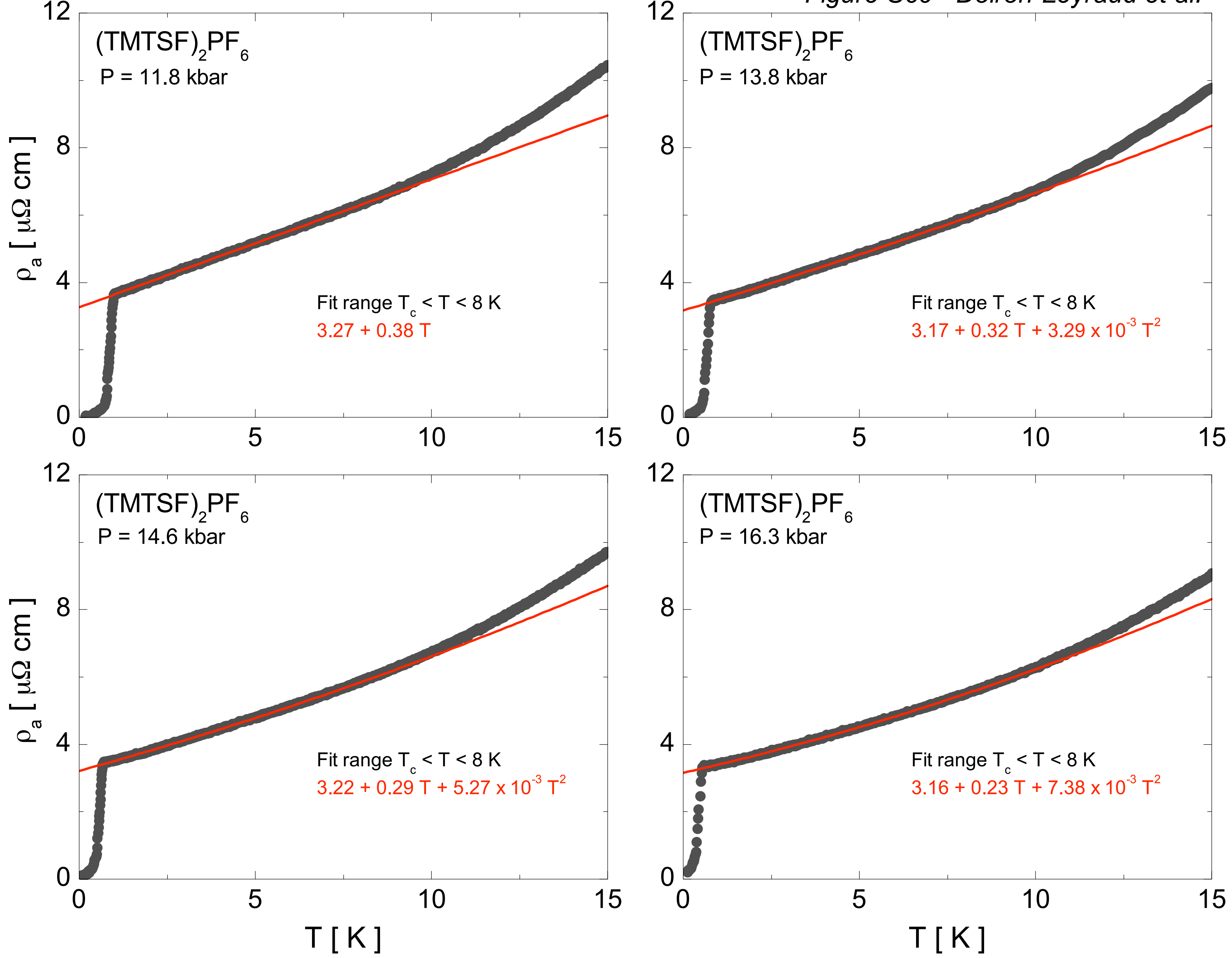

*Figure S6d - Doiron-Leyraud et al.*

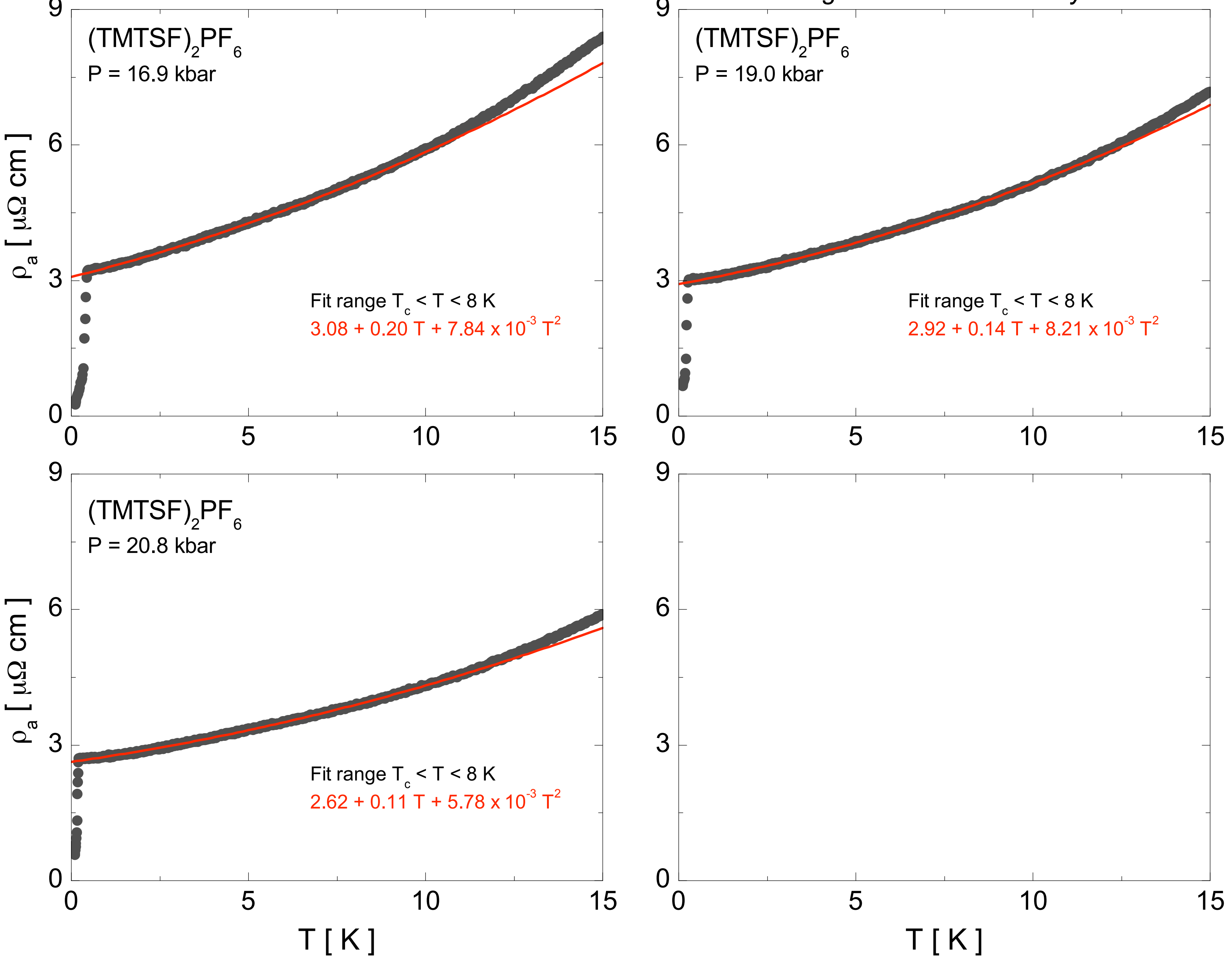

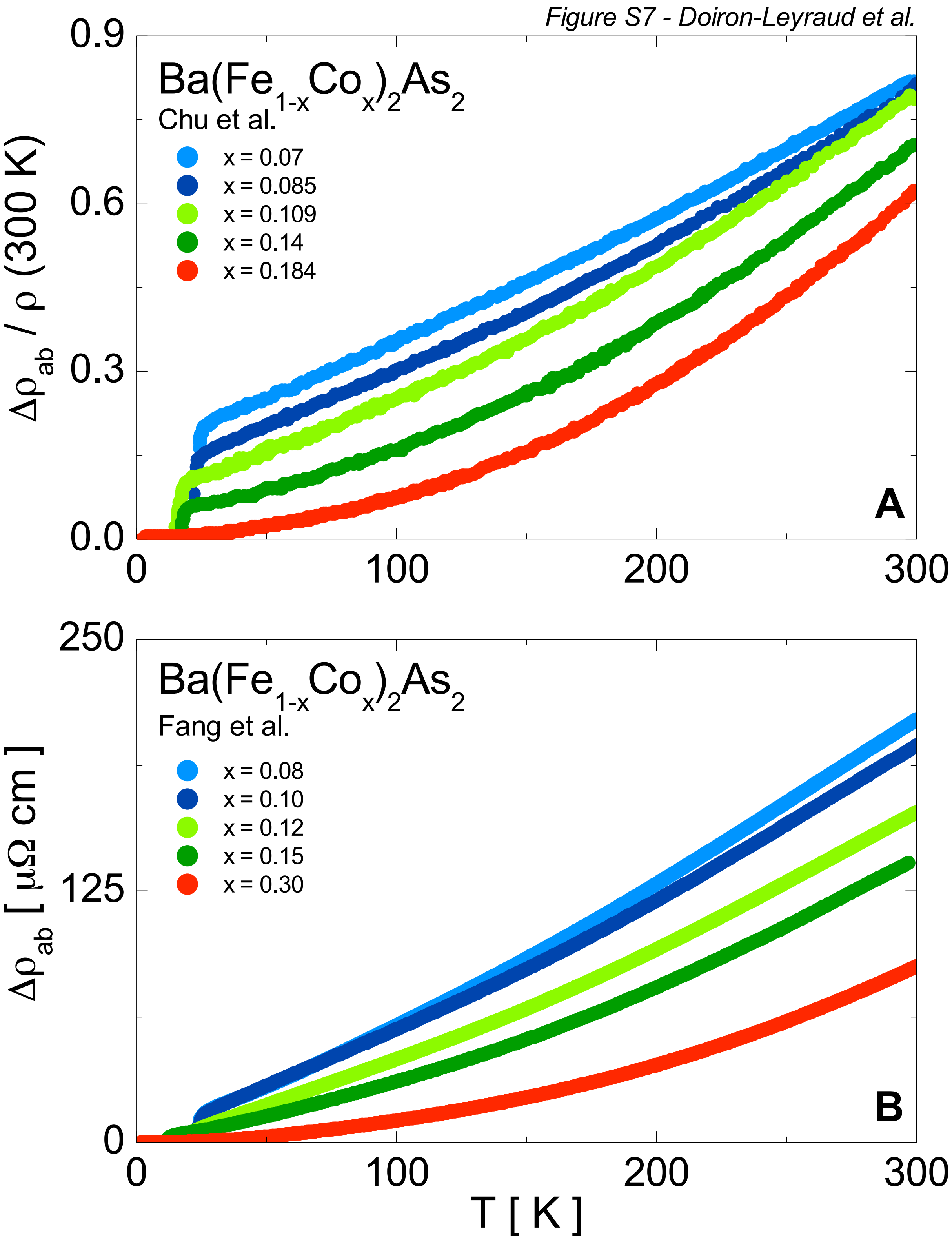

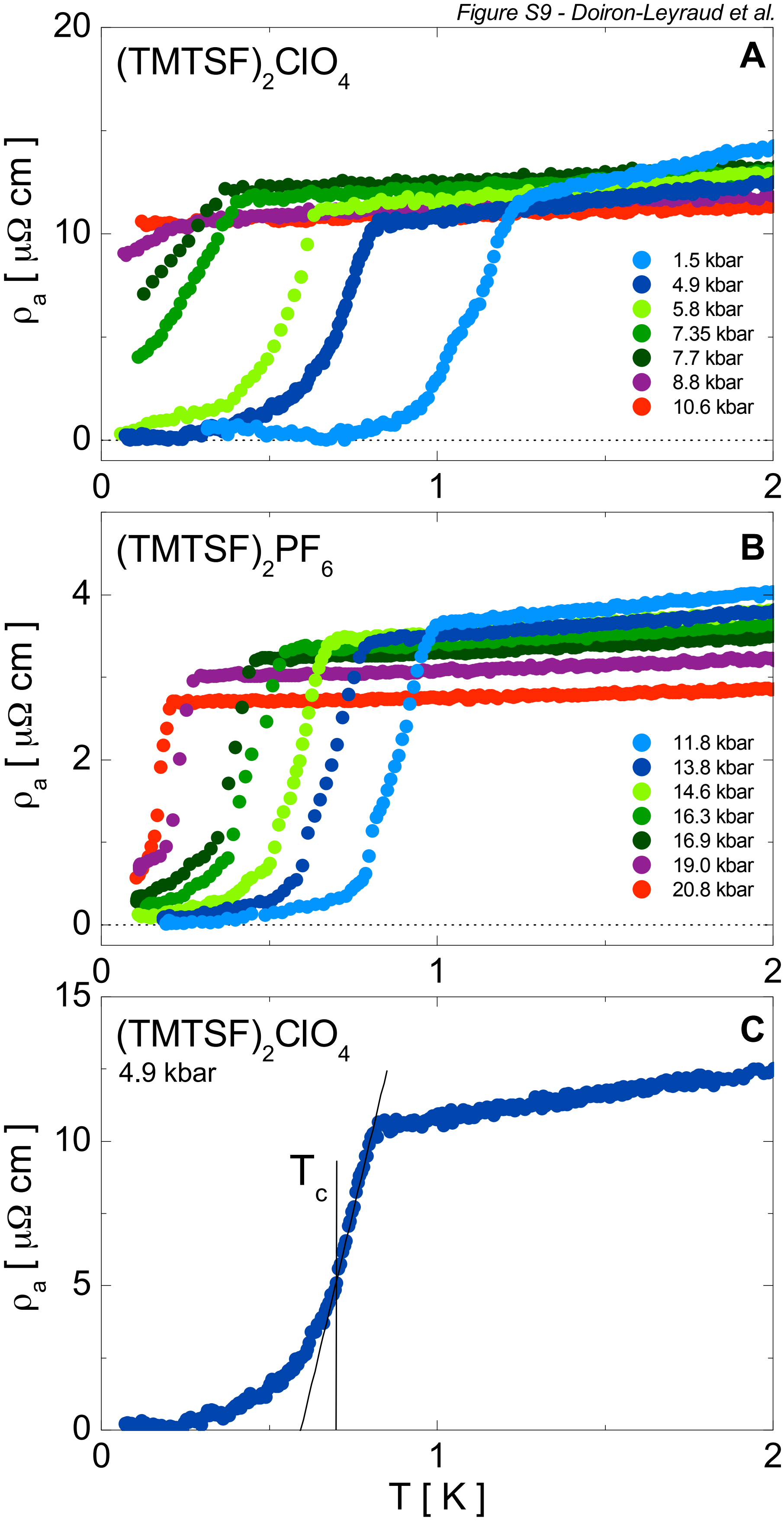

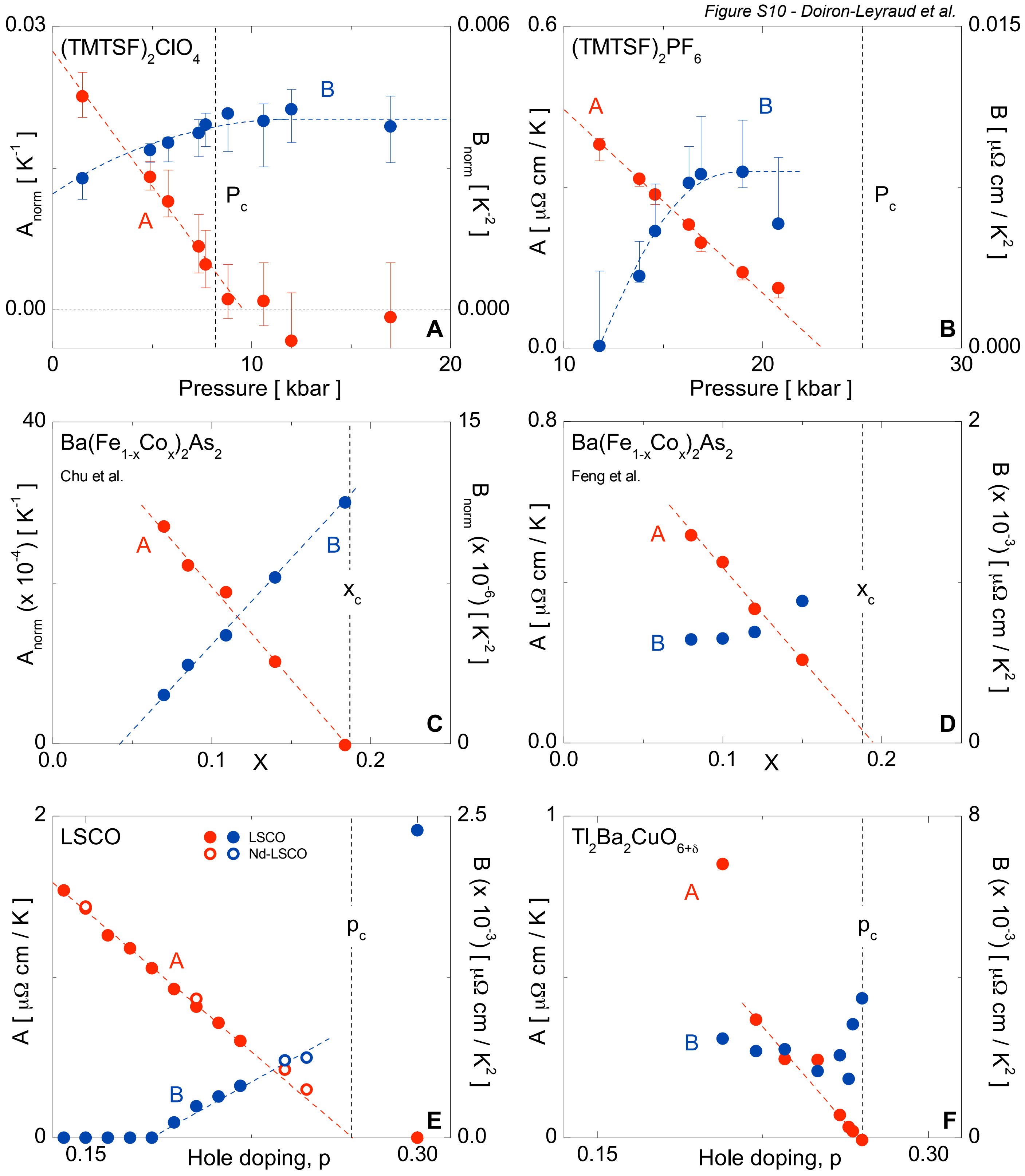

Figure S10 - Doiron-Leyraud et al.
(TMTSF)2ClO4
A
B
Pc
Anorm [ K-1 ]
Bnorm [ K-2 ]
Pressure [ kbar ]
(TMTSF)2PF6
A [ μΩ cm / K ]
B [ μΩ cm / K2 ]
Ba(Fe1-xCox)2As2
Chu et al.
Anorm (x 10-4) [ K-1 ]
Bnorm (x 10-6) [ K-2 ]
xc
C
Feng et al.
B (x 10-3) [ μΩ cm / K2 ]
X
D
LSCO
Nd-LSCO
pc
Hole doping, p
E
Tl2Ba2CuO6+δ
F